\documentclass[11pt]{article}
\usepackage[left=2cm,right=2cm,top=1cm,bottom=1.5cm]{geometry} % marges
\usepackage{amsmath}  	% pr le N de la loi normale
\usepackage{amssymb}  	% lettres mathématiques
\usepackage{amsfonts}
\usepackage{dsfont}
\usepackage{ulem}
\usepackage{pdfpages} 	% pr insérer un document PDF
\usepackage{setspace} % interlignes
\usepackage{caption} % legende des tables et figures
\usepackage{enumitem} % listes
\usepackage{authblk} % affiliations auteurs
\usepackage{endfloat}

\newcommand{\RVW}[1]{\textcolor{black}{#1}} % review

\begin{document}

%%%%%%%%%%%%%%%%%%%
% TITLE & AUTHORS %
%%%%%%%%%%%%%%%%%%%

\title{Joint models for the longitudinal analysis of measurement scales in the presence of informative dropout}

\author[,1]{Tiphaine SAULNIER*}
\author[1]{Viviane PHILIPPS}
\author[2,3]{Wassilios G MEISSNER}
\author[4,6]{Olivier RASCOL}
\author[4,5]{Anne PAVY-LE-TRAON}
\author[1,2,3]{Alexandra FOUBERT-SAMIER}
\author[1]{Cécile PROUST-LIMA}

\affil[1]{Univ. Bordeaux, Inserm, Bordeaux Population Health Research Center, UMR1219; F-33000 Bordeaux, France}
\affil[2]{CHU Bordeaux, Service de Neurologie des Maladies Neurodégénératives, IMNc, CRMR AMS; F-33000 Bordeaux, France}
\affil[3]{Univ. Bordeaux, CNRS, IMN, UMR 5293; F-33000 Bordeaux, France}
\affil[4]{French Reference Centre for MSA, University Hospital Toulouse; F-31000 Toulouse, France}
\affil[5]{Institut des Maladies Métaboliques et Cardiovasculaires, Inserm U1297, univ. Toulouse; F-31000 Toulouse, France}
\affil[6]{Inserm, Toulouse University and CHU Toulouse, Clinical Investigation Center CIC 1436, NS-Park/F-CRIN Network, NeuroToul COEN Center, and Departments of Neurosciences and Clinical Pharmacology; F-31000 Toulouse, France}

\date{}

\maketitle

\noindent * \textbf{Corresponding author } \\ 
Tiphaine SAULNIER \\
146 rue Léo Saignat, F-33000 Bordeaux, France \\
e-mail : tiphaine.saulnier@u-bordeaux.fr

\newpage

%%%%%%%%%%%%
% ABSTRACT %
%%%%%%%%%%%%
 
\section*{{Abstract}}
In health cohort studies, repeated measures of markers are often used to describe the natural history of a disease. Joint models allow to study their evolution by taking into account the possible informative dropout usually due to clinical events. However, joint modeling developments mostly focused on continuous Gaussian markers while, in an increasing number of studies, the actual \RVW{quantity} of interest is non-directly measurable; it \RVW{constitutes} a latent \RVW{variable} evaluated by a set of observed indicators from questionnaires or measurement scales. Classical examples include anxiety, fatigue, cognition. 
In this work, we explain how joint models can be extended to the framework of a latent quantity measured over time by \RVW{indicators} of different nature (e.g. continuous, binary, ordinal). The longitudinal submodel describes the evolution over time of the quantity of interest defined as a latent process in a structural mixed model, and links the latent process to each \RVW{observation of the indicators} through appropriate measurement models. Simultaneously, the risk of multi-cause event is modelled via a proportional cause-specific hazard model that includes a function of the mixed model elements as linear predictor to take into account the association between the latent process and the risk of event. Estimation, carried out in the maximum likelihood framework and implemented in the R-package JLPM, has been validated by simulations.
The methodology is illustrated in the French cohort on Multiple-System Atrophy (MSA), a rare and fatal neurodegenerative disease, with the study of dysphagia progression over time \RVW{stopped} by the occurrence of death.

\section*{Keywords}
joint model, time-to-event data, longitudinal data, latent process, ordinal data

\newpage

\section*{Highlights}
\begin{itemize}[label=\textbullet]
    \item Joint models (JM) are a useful tool to analyze longitudinal data with informative dropout
    \item Joint Latent Process Models (LPM) \RVW{extend JM to markers (of possibly different nature) measuring the same underlying quantity}
    \item Maximum Likelihood Estimation of the joint LPM is available in the R-package JLPM
    \item The case study quantifies the progression of dysphagia during M\RVW{ultiple-}S\RVW{ystem} A\RVW{trophy (MSA)} and its association with death risk
\end{itemize}

\newpage

\section*{Abbreviations}
\begin{description}
%    \item[CD4] Cluster of Differentiation 4
    \item[CI] Confidence Intervals
 %   \item[DIF] Differential Item Functioning
    \item[GRM] Graded Response Model
 %   \item[HIV] Human Immunodeficiency Virus
    \item[HR] Hazard Ratio
    \item[IRT] Item Response Theory
    \item[JM] Joint Model
    \item[JLPM] Joint Latent Process Model
    \item[MAR] Missing-At-Random
    \item[MNAR] Missing-Not-At-Random
    \item[MRI] Magnetic Resonance Imaging
    \item[MSA] Multiple-System Atrophy (disease)
    \item[MSA-C] MSA with predominant Cerebellar impairment
    \item[MSA-P] MSA with predominant Parkinsonism
%    \item[PSA] Prostate Specific Antigen
%    \item[RS] Response Shift
    \item[SD] Standard Deviation
    \item[UMSARS] Unified Multiple-System Atrophy Rating Scale
\end{description}

%%%%%%%%%%%%%%%%%%
%% MAIN ARTICLE %%
%%%%%%%%%%%%%%%%%%

\newpage

%----------------------------------------------%
%--               INTRODUCTION               --%
%----------------------------------------------%

\section{Introduction}

In health cohort studies, markers measured at multiple times are often used to describe the natural history of a disease, monitor patients or predict the risk of clinical progression. Classical examples include T-cell CD4 counts and viral load for the progression of HIV \cite{tsiatis_joint_2004} or PSA for prostate cancer evolution \cite{ferrer_joint_2016}. Due to the intrinsic intra-subject correlation between the repeated measures of markers, their evolution can not be modelled using classical regression models, and mixed models which include individual random effects to account for this serial correlation are now adopted worldwide \cite{commenges_dynamical_2015}.

During the follow-up, clinical events (e.g. diagnosis, recurrence, death) may disrupt the progression of the markers and induce a dropout. When this interruption %truncation 
of the follow-up is predictable by the observed marker data, the missing data mechanism is called missing-at-random (MAR), and inference provided by the mixed model is still valid \cite{thomadakis_longitudinal_2019}. However, in many cases, the dropout is likely to depend on the underlying (unobserved) disease mechanism rather than only on the strictly observed data. The missing data mechanism becomes missing-not-at-random (MNAR), and mixed models may not provide correct inference anymore \cite{little_modeling_1995,rouanet_interpretation_2019}. 
During the last twenty years, the statistical community has massively embraced the issue of dropout in longitudinal analysis which lead to the development of joint models for the simultaneous analysis of repeated markers and clinical events \cite{rizopoulos_joint_2012,henderson_joint_2000,tsiatis_joint_2004,proust-lima_joint_2014}. Joint models combine a mixed model describing the progression of the markers and a survival model for the time of occurrence of the clinical events, while associating the two models through shared random variables, usually the random effects from the mixed model \cite{rizopoulos_joint_2012}. 
Beyond the study of the marker progression in the presence of MNAR dropout, this method also more generally assesses the association between a marker trajectory and an associated event of interest \cite{he_joint_2016,wang_joint_2019,ferrer_joint_2016}.

Despite many developments in joint models in the recent years \cite{alsefri_bayesian_2020,hickey_joint_2018}, most works are dedicated to classical continuous biomarkers stemmed from blood samples, MRI, etc. Yet, in an ever-increasing number of health studies, \RVW{the actual quantity of interest is not directly measured. It is a latent construct which is assessed using a set of indicators measured with error, usually stemmed from questionnaires or measurement scales}. Examples include health related quality of life in Cancer research and beyond \cite{fayers_quality_2002}, cognitive functioning and functional dependency in neurodegenerative diseases \cite{edjolo_natural_2016,proust-lima_analysis_2013}, or many other constructs such as anxiety or depressive symptomatology \cite{james_exploring_2020,kruyen_shortening_2013}. The analysis of latent constructs stemmed from measurement scales and questionnaires requires a specific attention. Measurement scales usually translate into multiple categorical and/or continuous items that measure different aspects of the underlying construct of interest. Furthermore, when aggregating the item information into a sumscore, the resulting univariate marker may not have classical Gaussian properties for continuous markers; they are usually bounded with floor/ceiling effects and possible unequal-interval scaling \cite{proust-lima_analysis_2013,proust-lima_are_2019}. 

In this contribution, we show how the joint modeling methodology can be extended to handle repeated data from measurement scales in the presence of an informative clinical event, and provide a software solution with the R package JLPM. We illustrate the methodology through simulations and in Multiple-System Atrophy (MSA), a rare and deadly neurodegenerative disease, which progression is almost exclusively described by measurement scales \cite{foubert-samier_disease_2020}. 

\RVW{The article is organized as follows: Section 2 presents the methodology developed to handle previously mentioned challenges (multiple markers of different nature, competing risks, delayed entry), Section 3 reports simulation studies aiming at validating the inference procedure. Section 4 illustrates the methodology with the analysis of dysphagia progression in MSA, and Section 5 concludes.}

%---------------------------------------------%
%--               METHODOLOGY               --%
%---------------------------------------------%

\section{Methodology}

Let consider a cohort of $N$ individuals followed up over time.
We define $T^*_i$ the time of occurrence of an event of interest \RVW{for subject $i$ ($i=1,...,N$)}. This event may be due to $P$ different causes, defining a competing risk setting. As some individuals may be censored before this occurrence, we observe $T_i = min(T^*_i, C_i)$ where $C_i$ is the censoring time, and $\delta_i$ the event indicator such that $\delta_i=p$ when the event of cause $p$ occurs first and before censoring ($T^*_i \leq C_i$), and $\delta_i=0$ otherwise. We also collect repeated measures of $K \geq 1$ markers measuring the same construct of interest (for instance $K$ items of a measurement scale or only 1 isolated item, or 1 sumscore). The marker values are noted $Y_{ikj}$ for subject $i$ ($i=1,...,N$) and marker $k$ ($k=1,...,K$) at time $t_{ikj}$ with $j$ the repeated occasion ($j=1,...,n_{ik}$). The event of interest interrupts the collection of the repeated markers so that necessarily $t_{ikn_{ik}} \leq T_i$. In this methodology, the markers do not need to be measured at the same visit times across markers and across individuals.

\begin{figure}[h!]
\caption{Schematic diagram of the joint model structure for $K$ repeated markers measuring the same construct and a time of event (possibly of multiple causes)}
\centering
\includegraphics[width = 15cm, trim= 0cm 5.5cm 15cm 1cm, clip, page = 1]{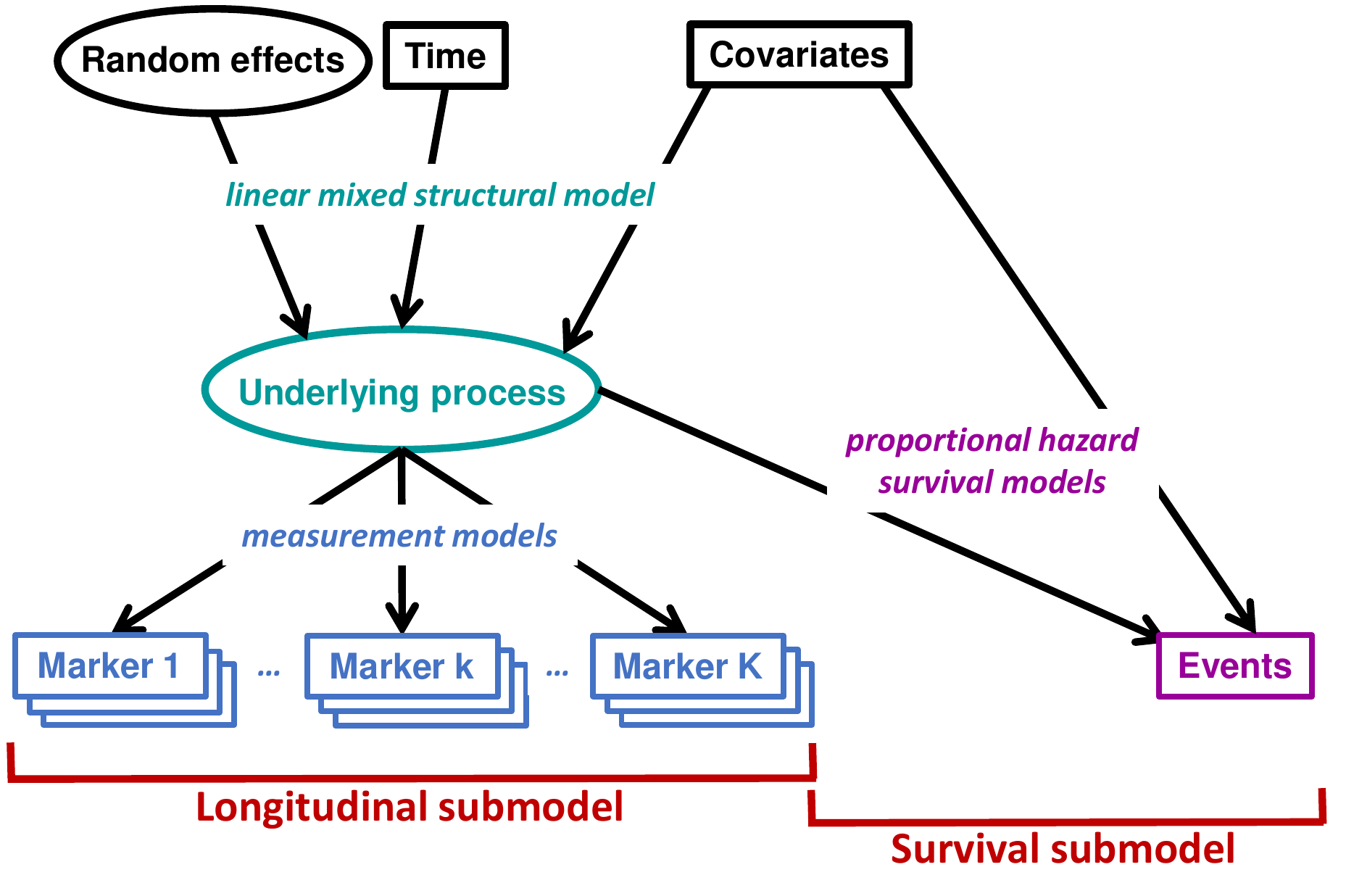}
\label{fig:schema_methodo}
\end{figure}

The joint model aims to analyze the trajectory of the construct of interest over time and the risk of event by accounting for their correlation. As shown in Figure \ref{fig:schema_methodo}, the joint model includes two submodels, one for the longitudinal process (on the left) and one for the time-to-event process (on the right). They are detailed below.

%~~~~~~~~~~~~~~~~~~~~~~~~~~%
%~   longitudinal model   ~%
%~~~~~~~~~~~~~~~~~~~~~~~~~~%

\subsection{Longitudinal model}

When analyzing data from measurement scales or questionnaires, and more generally psychometric data, one makes the distinction between the construct of interest which is a latent process, denoted $\Delta_i(t)$ and defined in continuous time ($t \in \mathbb{R}$), and its observations, denoted $Y_{ikj}$, measured \RVW{with error} at discrete time visits $t_{ikj}$. As emphasized in Figure \ref{fig:schema_methodo}, the longitudinal model thus consists of a structural model describing the latent process over time, and measurement models defining its link with each marker \cite{proust-lima_analysis_2013}. 

%------------------------%
%-   structural model   -%
%------------------------%

\subsubsection{Structural model} \label{sec:structuralmodel}

The trajectory over time of $\Delta$ can be described using a linear mixed model \cite{laird_random-effects_1982} as follows :

\begin{equation} \label{eq:Delta}
	\Delta_i(t) = X_i(t)^{L\top}\beta^L + Z_i(t)^{\top}b_i
\end{equation}

with $X^L_i(t)$ and $Z_i(t)$ two vectors of covariates associated to $\beta^L$ the vector of fixed effect parameters, and $b_i$ the $q$-vector of normally distributed individual random effects ($b_i \sim \mathcal{N}(0, B)$), respectively. We note that, as the quantity described here is not directly observed or measured, we do not consider any measurement error in the equation. The term $X_i(t)^{L\top}\beta^L$ defines the mean trajectory of the latent process at the population level while $Z_i(t)^{\top}b_i$ captures the individual deviation to the mean trajectory.

%-------------------------%
%-   measurement model   -%
%-------------------------%

\subsubsection{Measurement model}

The measurement models define the nature of the link between the latent process $\Delta$ and each marker, considered as a noisy measure of $\Delta$, \RVW{that is a marker of $\Delta$ prone to measurement error}. Depending on the study to be conducted and the data available, the markers may vary in number and nature (e.g., continuous, ordinal). We review here different measurement models according to the nature of the markers. In any case, one measurement model is to be defined for each marker $k \in \{1,...,K\}$. \\

%------------------%
%-   continuous   -%
%------------------%

%----------------%
%-   gaussian   -%
%----------------%

\textbf{Continuous Gaussian marker} 

Classically, continuous markers are considered as having a multivariate normal distribution with Gaussian errors. In this case, the measurement model for marker $k$ is: 

\begin{equation} \label{eq:mksG}
	Y_{ik}(t_{ikj}) = \eta_{k1} + \eta_{k2}\times(\Delta_i(t_{ikj}) + \epsilon_{ikj})
\end{equation}

where $\eta_{k1}$ and $\eta_{k2}$ are two marker-$k$-specific parameters transforming the continuous marker into the latent process scale, and $\epsilon$ is the independent Gaussian measurement error ($\epsilon_{ikj} \sim \mathcal{N}(0, \sigma_k^2)$). \\

In the case of one single Gaussian marker ($K=1$), $\eta_{k1}=0$ and $\eta_{k2}=1$ reduces the equation to the standard definition of a linear mixed model:

\begin{equation} \label{eq:1mkG}
	Y_{i}(t_{ij}) = \Delta_i(t_{ij}) + \epsilon_{ij} = X_i(t_{ij})^{L\top}\beta^L + Z_i(t_{ij})^{\top}b_i + \epsilon_{ij} 
\end{equation}

%--------------------%
%-   non-gaussian   -%
%--------------------%

\textbf{Continuous non-Gaussian marker}

When analyzing data from measurement scales, the sum-score of all or a part of the items is frequently considered as a relevant marker of the underlying process of interest. This score is often treated as continuous because of its large number of possible levels. However its distribution is very rarely Gaussian due to ceiling and floor effects induced by its lower and upper bounds, and due to the fact that, as based on the sum of ordinal items, a one-point change does not necessarily have the same meaning depending on the level of the score. This \RVW{phenomenon}, called curvilinearity, is not compatible with the assumptions of the linear model and linear mixed model \cite{proust-lima_misuse_2011,proust-lima_are_2019}. 
In this context, a continuous but non-gaussian marker $k$ can be described using a curvilinear measurement model where a parameterized transformation function $H_k$ normalizes marker $k$:

\begin{equation} \label{eq:mks-C}
	H_k( Y_{ik}(t_{ikj}); \eta_k ) = \Delta_i(t_{ikj}) + \epsilon_{ikj}
\end{equation}

with $\epsilon_{ikj} \sim \mathcal{N}(0, \sigma_k^2)$ and $H_k(.;\eta_k)$ the marker-$k$-specific parametric link function. This function is monotonically increasing and is often approximated by splines \cite{proust-lima_analysis_2013}. We can notice that if $H_k(.)$ is a linear function, this model turns into the classical linear mixed model defined in the previous subsection. \\

%---------------%
%-  discrete   -%
%---------------%

\textbf{Discrete marker}

\RVW{Ordinal items (including binary items as 2-level ordinal items)} are frequently considered for measuring complex constructs. They can be described using proportional odds logistic models or cumulative probit models \cite{commenges_dynamical_2015,proust-lima_continuous-time_2021}. Such models are based on the assumption of increasing monotonicity indicating that a higher level of the item reflects a higher degradation of the latent dimension that it measures. This is formalized as follows:

\begin{equation} \label{eq:mks-D}
\begin{gathered}
	Y_{ik}(t_{ikj}) = m \Leftrightarrow \eta_{k, m} < \Delta_i(t_{ikj}) + \epsilon_{ikj} \leq \eta_{k, m+1} \textrm{ with $m$} \in \{0, 1, ..., M_k\} \\ \textrm{ and } -\infty= \eta_{k,0} \leq \eta_{k, 1} \leq ... \leq \eta_{k, m} \leq \eta_{k, m+1} \leq ... \leq \eta_{k, M_k+1}= +\infty
\end{gathered}
\end{equation}

where $\epsilon_{ikj}$ is an additional noise either Logistic in the proportional odds logistic model or Gaussian in the cumulative probit  \cite{commenges_dynamical_2015}. We consider here the latter with $\epsilon_{ikj} \sim \mathcal{N}(0,\sigma_{k}^2)$. 

Parameters $\eta_k=(\eta_{k,m})_{m=1,...,M_k}$ are to be estimated. Also called thresholds or locations, they correspond to the level of the latent construct at which the probability of observing $Y_{ik}(t_{ikj})$ lower/higher $m+1$ is 0.5. These measurement models are usual in \textit{Item Response Theory} (IRT) where they are called Graded Response Models (GRM) \cite{baker_item_2004}.

\subsubsection{Caution note}

In the case of multiple markers, this framework relies on important assumptions: 
\begin{itemize}
\item unidimensionnality of the latent construct. The model assumes that all markers should measure the same phenomenon, the latent dimension of interest. In the case of a multidimensional measurement scale, items should be grouped by unidimensional underlying dimensions, and different dimensions analyzed separately. 
\item inter-marker and intra-marker conditional independence given the latent construct. The latent construct should constitute the only source of shared information across pairs of markers and within repeated measures of the same marker. In the case of residual dependency across markers, redundant markers may be deleted or their correlation modeled in the measurement models. In the case of residual dependency within marker, additional random-effect specific to the marker and/or marker-specific association with covariate may be added \cite{proust-lima_analysis_2013}. 
\end{itemize}
These assumptions \RVW{need to be} verified in preliminary analyses to prevent any bias in the results. \RVW{For this purpose, we recommend to follow the PROMIS strategy \cite{reeve_psychometric_2007}}. Note that intra-marker conditional independence given the latent construct may be assessed in sensitivity analyses.

\subsubsection{Identifiability Constraints}
Except for the specific case of equation \eqref{eq:1mkG}, the longitudinal submodel, as defined in this section, is not identifiable. Due to the introduction of a latent process, identifiability constraints must be added \RVW{to determine the dimension of the latent process; this is a requirement in any latent variable model.} In this work, we chose to center and reduce the latent process $\Delta$ at time 0 in the reference category. This translates into an intercept fixed to $\beta^L_{\text{int}}=0$ (location constraint) and a variance of the first random-effect (usually the random intercept) fixed to $B_{1,1}=1$ (dispersion constraint) \cite{commenges_dynamical_2015}. This implies that the unit of the latent process corresponds to the residual inter-individual standard deviation of the latent quantity when time $t=0$.

%~~~~~~~~~~~~~~~~~~~~~~%
%~   survival model   ~%
%~~~~~~~~~~~~~~~~~~~~~~%

\subsection{Survival model}

The survival model describes the risk of occurrence of each cause of event, usually using a cause-specific proportional hazard model \cite{commenges_dynamical_2015}. In shared random effect joint models, the instantaneous risk function depends on a (possibly time-dependent) function of the elements of the structural model \eqref{eq:Delta} noted $g_{ip}(b_i,t)$. This element, which notably depends on the random effects $b_i$, is added as a linear predictor in the survival model to quantify the intensity of the association between the latent dimension and the considered event $p$. 

The cause-specific hazard model is described in continuous time $t$ ($\in \mathbb{R}$) for each cause $p=1,...,P$ and each subject $i$ ($\in \{1,...,N\}$) as follows:

\begin{equation} \label{eq:surv}
	%\lambda_{ip}(t) = \lambda_{0p}(t;\psi_p) \exp ( X^{\top}_{spi}\beta_{sp} + g_{ip}(b_i, t)^{\top}\eta_p )
	\lambda_{ip}(t) = \lambda_{0p}(t;\psi_p) \exp ( X^{S\top}_{pi}\beta^S_{p} + g_{ip}(b_i, t)^{\top}\alpha_p )
\end{equation}

with $\lambda_{0p}(.)$ the parametric baseline risk function associated to the vector of parameters $\psi_p$ for cause $p$. This function is usually chosen among Weibull hazards, piecewise constant hazards or approximated using spline functions. The vector of exogenous time-independent covariates $X^S_{p}$ is associated with the vector of parameters $\beta^S_{p}$, and the function $g_{ip}(b_i,t)$ is associated with the vector of parameters $\alpha_p$ which quantifies the association between the two processes. Many different functions $g_{ip}(b_i,t)$ can be specified \cite{sene_shared_2013}. We focus here on \RVW{two frequently used} structures \RVW{encountered in the joint modeling literature \cite{rizopoulos_joint_2012,henderson_joint_2000} }:

\begin{itemize}
    
    \item[-] \textbf{the vector of random effects (dimension $q$)}: the survival model is adjusted on the individual deviations to the mean latent process trajectory with
        \begin{equation} \label{eq:surv_RE}
	        g_{ip}(b_i, t) = g_{ip}(b_i) = b_i
        \end{equation}
    
    \item[-] \textbf{the current latent process level (dimension $1$)}: the instantaneous risk is adjusted on the level of the underlying process (defined in Equation \eqref{eq:Delta}) at the same time with
        \begin{equation} \label{eq:surv_CL}
	        g_{ip}(b_i, t) = \Delta_i(t)
        \end{equation}
    
\end{itemize}

%~~~~~~~~~~~~~~~~~~~~~~~%
%~   estimation part   ~%
%~~~~~~~~~~~~~~~~~~~~~~~%

\subsection{Estimation}

Let $\Theta$ denote the vector of all sub-model parameters including :
\begin{itemize}
    \item[-] for the structural model: vector $\beta^L$ of fixed effects, parameters of the random-effects variance-covariance matrix $B$;
    \item[-] for the measurement model: for all marker $k \in \{1,...,K\}$, variance $\sigma^2_k$ of the measurement error and vector $\eta_k$ of parameters, it can be regression parameters $\eta_k = (\eta_{k1},\eta_{k2})$ if marker $k$ is Gaussian, link function parameters $\eta_k$ if marker $k$ is continuous but not necessarily Gaussian, or thresholds $\eta_k = (\eta_{k,m})_{m \in \{1,...,M_k\}}$ if marker $k$ is ordinal;
    \item[-] for the survival model: for all cause $p \in \{1,...,P\}$, vector $\psi_p$ of parameters in the baseline hazard function, vector of parameters $\beta^S_{p}$ for all covariates included in the survival model, and $\alpha_p$ the association parameters.
\end{itemize}
The joint model parameters $\Theta$ can be estimated by maximizing the log-likelihood $ \displaystyle \mathcal{L}(\Theta) = \sum_{i=1}^{N} \log L_i(\Theta)$ where $L_i(\Theta)$ is the individual contribution to the likelihood. 

\subsubsection{Individual contribution to the likelihood}

Thanks to the assumption of independence between the latent dimension $\Delta_i$, measured by the markers $Y_i = (Y_{ik})_{k=1,...,K}$, and the time-to-event $(T_i,\delta_i)$ \RVW{conditionally} to the random effects $b_i$, the individual contribution to the likelihood for subject $i \in \{1,...,N\}$ can be developed as follows:

\begin{equation} \label{eq:indlikelihood}
	\begin{array}{ll}
		L_i(\Theta)& = f_{Y_i, (T_i, \delta_i)}(Y_i, (T_i, \delta_i);\Theta) \\
		% & = \int_{\mathbb{R}^q} f_{Y_i | b_i}(Y_i | b;\Theta) f_{(T_i, \delta_i) | b_i}((T_i, \delta_i) | b;\Theta) f_{b_i}(b; \Theta) db \\
		& = \int_{\mathbb{R}^q} \prod_{k=1}^{K} \{ f_{Y_{ik} | b_i}(Y_{ik} | b;\Theta) \} f_{(T_i, \delta_i) | b_i}((T_i, \delta_i) | b;\Theta) f_{b_i}(b; \Theta) db
	\end{array}
\end{equation}
with $f(.)$ the generic notation for a density function.

The density of the repeated marker depends on its nature. The density of a continuous marker can be expressed using the change of variable theorem:

\begin{equation} \label{eq:indlikelihood1_c}
	\begin{array}{ll}
	   \displaystyle
	    f_{Y_{ik} | b_i}(Y_{ik} | b;\Theta) = \phi(\tilde{Y}_{ik} | b ;\Theta) \times \prod_{j=1}^{n_{ik}} J(\tilde{Y}_{ikj}|b)
	\end{array}
\end{equation}
where $\tilde{Y}_k = H^{-1}_k(Y_k)$ designates the value of marker $Y_k$ transformed by its link function. Functions $\phi(.)$ and $J(.)$ are the density function of a normal distribution and the jacobian associated to the link function, respectively.

For a discrete marker, it can be expressed as follows:
\begin{equation} \label{eq:indlikelihood1_d}
	\begin{array}{ll}
	   \displaystyle
	    f_{Y_{ik} | b_i}(Y_{ik} | b;\Theta) & = \prod_{j=1}^{n_{ik}} \prod_{m=0}^{M_k} \mathbb{P}(Y_{ik}(t_{ikj}) = m | b)^{\mathds{1}_{Y_{ik}(t_{ikj}) = m}} \\
	    & = \prod_{j=1}^{n_{ik}} \prod_{m=0}^{M_k} [ \Phi( \frac{1}{\sigma_k}(\eta_{k,m+1} - \Delta_i(t_{ikj}))) - \Phi( \frac{1}{\sigma_k}(\eta_{k,m} - \Delta_i(t_{ikj}))) ]^{\mathds{1}_{Y_{ik}(t_{ikj}) = m}}
	\end{array}
\end{equation}
where $\Phi(.)$ is the standard normal distribution function.

The density of the time-to-event is:

\begin{equation} \label{eq:indlikelihood2}
	\begin{array}{ll}
	    \displaystyle 
	    f_{(T_i, \delta_i) | b_i}((T_i, \delta_i) | b;\Theta) = S_i(T_i | b;\Theta) \prod_{p=1}^{p} \lambda_{ip}(T_i | b;\Theta) ^{\mathds{1}_{\delta_i=p}}
	\end{array}
\end{equation}
with $\lambda_{ip}(.)$ the hazard function specific to cause $p$ given in Equation \eqref{eq:surv} and $S_i(.)$ the survival function combining all $P$ causes:

\begin{equation} \label{eq:indlikelihood2bis}
	\begin{array}{ll}
	    \displaystyle 
	    S_i(t | b; \Theta) = \prod_{p=1}^{P} S_{ip}(t | b; \Theta)
	    = \prod_{p=1}^{P} \exp ( - \int_0^t \lambda_{ip}(s | b; \Theta) ds )
	\end{array}
\end{equation}

Finally, the density of the random effects is the density of a centered multivariate normal distribution: 

\begin{equation} \label{eq:indlikelihood3}
	\begin{array}{ll}
	    f_{b_i}(b; \Theta) = \phi_{\mathcal{N}(O,B)}(b)
	\end{array}
\end{equation}

In the presence of delayed entry, that is when a subject is included in the study some time after time zero, the time-to-event is left-truncated. This is accounted for by considering instead the individual contribution to the likelihood for truncated data $\displaystyle L_{trunc,i}(\Theta)$ in which the naive individual contribution $L_i(\Theta)$ is divided by the probability to survive until the entry time $T_{0i}$: 

\begin{equation} \label{eq:delayedentry}
	\begin{array}{ll}
	    \displaystyle L_{trunc,i}(\Theta) = \frac{L_i(\Theta)}{S_i(T_{0i;\Theta})} 
	\end{array}
\end{equation}

\subsubsection{Software}

\RVW{Maximum likelihood estimation of the joint model parameters} is implemented \RVW{in} the R package \textbf{JLPM} (https://github.com/VivianePhilipps/JLPM). The optimization of the log-likelihood is numerically carried out by a Marquardt-Levenberg algorithm with stringent convergence criteria based on the first and second derivatives of the log-likelihood (see \cite{philipps_robust_2021} for details).  Example scripts are given in the package. \RVW{To reduce the computation time, optimization can be carried out in parallel mode (i.e., using multiple cores).}

\RVW{As with any complex model, we recommend to proceed step-by-step by estimating submodels with increasing complexity (e.g., survival submodel without dependency on the latent process, longitudinal model with a reduced number of random effects) to determine plausible initial values, and as such, reduce the number of iterations of the optimization algorithm applied to the final model}. 

The integral on the $q$ random effects in Equation \eqref{eq:indlikelihood} does not have analytical solution so that the computation of the likelihood involves numerical integrations. This can be carried out with a Monte-Carlo algorithm which consists in simulating many draws of the random effects (generally around 100000) and to compute and then average the results of the function to integrate. In this package, \RVW{we used instead a} quasi Monte-Carlo algorithm \cite{pan_quasi-monte_2007}. This integration method uses a deterministic sequence rather than a random one. Then, smaller integration errors are obtained due to the low discrepancy of the chosen deterministic sequence, so that the number of required points is usually reduced to 1000. 

Moreover, when the association structure $g(b_i,t)$ is time-dependent (not the case for the dependence on the random effects), the survival function in Equation \eqref{eq:indlikelihood2bis} also involves an integral with no analytical solution; it is approximated by a  Gauss-Kronrod gaussian quadrature with 15 points \cite{gonnet_review_2012}.

\section{Simulation studies}

\RVW{Two simulation studies were performed to illustrate the methodology and validate the estimation procedure implemented in the R-package \textbf{JLPM} according to the the type of markers and times-to-event included, and the nature of the dependency structure between the longitudinal and the survival submodels.}

\subsection{Simulation design}

\RVW{The first simulation study, described below, included 2 ordinal and 2 Gaussian markers measuring the same underlying process, 2 competing causes of event and a dependence on the shared random effects as in \eqref{eq:surv_RE}. The second one, reported only in supplementary materials, included 4 ordinal outcomes, 1 cause of event and a dependence on the current process level as in \eqref{eq:surv_CL}}. 

\RVW{We simulated 500 samples of 300 subjects each. The structural model for the underlying process consisted in a linear function of time at the population ($\beta^L = (0,1)^\top$) and individual level ($B = \big(\begin{smallmatrix} 1 & 0\\ 0 & 0.2 \end{smallmatrix}\big)$). For each cause of event $p$ ($p$ = $1$ or $2$), the time-to-event was defined by a Weibull baseline risk function $\lambda_{0p}(t;\psi_p)=\psi_{p1}^{\psi_{p2}}\psi_{p2}t^{\psi_{p2}-1}$ with $\psi_{11} = 0.2, \psi_{12} = 5$ for cause $1$ and $\psi_{21} = 0.198, \psi_{22} = 8$ for cause $2$. The unique linear predictor was the random effects of the underlying process with association parameter fixed to $\alpha_1 = (0.1,0.2)$ for cause $1$ and to $\alpha_2 = (0.3,0.2)$ for cause $2$. }

\RVW{Visit times were generated every year (or time unit) from year 0 and up to the minimum between year 4 (administrative censoring) and the time-to-event. A delayed entry was generated to translate an individual-specific time of entry (year 0). To do so, we generated individual time-of-entry from a continuous uniform distribution defined on the interval [0,2]. This lead to a time-of-entry mean around 1.00 (SD=0.58), 21.09\% of censoring on average and 3.64 repeated measures of each marker on average. The 2 discrete markers had 4 ordinal levels each. Those marker data were generated according to equation \eqref{eq:mks-D} with thresholds $\eta_1 = (0.5,1,1.5)^{\top}$ for marker $1$ and $\eta_2 = (0.25,0.75,0.8)^{\top}$ for marker $2$. The 2 Gaussian marker data were generated according to equation \eqref{eq:mksG} with parameters $\eta_3=(1,0.4)^{\top}$ for marker $3$ and $\eta_4=(2,0.2)^{\top}$ for marker $4$. Measurement error variances were fixed to $\sigma=1$ for all markers. }

\subsection{Results}

We report in Table \ref{tab:simu} the mean estimate, relative bias, variance estimate (empirical or asymptotic) and the coverage rate of the 95\% confidence interval for each parameter. 
The estimation procedure provides very good results on this example: for all parameters, \RVW{the bias is negligible}, the mean asymptotic variance is close to the empirical variance, and the coverage rate of the 95\% confidence interval is very close to the nominal value.

\begin{table}[p]
    \caption{Summary on 500 replicates \RVW{of the estimation of a joint model }with \RVW{4 markers (2 ordinal and 2 Gaussian) and 2 competing events} with shared random effect dependency structure \RVW{on} samples of 300 individuals.}
    \begin{center}
        \begin{tabular}{lcccccc}
            \hline
            %& & & & & mean &  \\
            parameters & true & mean &relative & empirical & mean asymptotic & 95\%CI \\ 
                        & value & estimate & bias & standard & standard & coverage \\
                        & & & (in \%) & deviation & deviation & rate (in \%) \\
            \hline
            \multicolumn{7}{l}{\emph{survival model}} \\
                \multicolumn{7}{l}{\hspace{0.5cm} baseline risk functions(Weibull)} \\
                    \multicolumn{7}{l}{\hspace{1cm} \textit{cause $1$}} \\
                        \hspace{1.5cm} scale $\sqrt{\psi_{11}}$ & 0.447 & 0.447 & 0.0 & 0.004 & 0.004 & 94.0 \\
                        \hspace{1.5cm} shape $\sqrt{\psi_{12}}$ & 2.236 & 2.249 & 0.6 & 0.087 & 0.085 & 94.8 \\
                    \multicolumn{7}{l}{\hspace{1cm} \textit{cause $2$}} \\
                        \hspace{1.5cm} scale $\sqrt{\psi_{21}}$ & 0.445 & 0.445 & 0.0 & 0.003 & 0.003 & 94.4 \\
                        \hspace{1.5cm} shape $\sqrt{\psi_{22}}$ & 2.828 & 2.856 & 1.0 & 0.109 & 0.112 & 95.0 \\
                \multicolumn{7}{l}{\hspace{0.5cm} association parameters} \\
                    \multicolumn{7}{l}{\hspace{1cm} \textit{cause $1$}} \\
                        \hspace{1.5cm} $\alpha_{1,intercept}$ & 0.100 & 0.112 & 12.0 & 0.111 & 0.116 & 96.0 \\
                        \hspace{1.5cm} $\alpha_{1,slope}$ & 0.200 & 0.183 & -8.4 & 0.276 & 0.258 & 93.4 \\
                    \multicolumn{7}{l}{\hspace{1cm} \textit{cause $2$}} \\
                        \hspace{1.5cm} $\alpha_{2,intercept}$ & 0.300 & 0.318 & 6.0 & 0.143 & 0.140 & 95.8 \\
                        \hspace{1.5cm} $\alpha_{2,slope}$ & 0.200 & 0.200 & 0.0 & 0.302 & 0.294 & 95.6 \\
            \multicolumn{7}{l}{\emph{structural model}} \\
                \multicolumn{7}{l}{\hspace{0.5cm} adjustment covariates, $\beta^L$} \\
                    \hspace{1cm} intercept & 0.000 & - & - & - & - & - \\
                    \hspace{1cm} time & 1.000 & 1.005 & 0.5 & 0.079 & 0.078 & 93.8 \\
                \multicolumn{7}{l}{\hspace{0.5cm} random effect covariance parameters, $B$} \\
                    \hspace{1cm} choleski 0 & 1.000 & - & - & - & - & - \\
                    \hspace{1cm} choleski 1 & 0.000 & 0.007 & - & 0.047 &  0.039 & 87.4 \\
                    \hspace{1cm} choleski 2 & 0.447 & 0.449 & 0.4 & 0.044 & 0.039 & 92.0 \\
            \multicolumn{7}{l}{\emph{outcome-specific measurement model}} \\
                \multicolumn{7}{l}{\hspace{0.5cm} thresholds for discrete markers} \\
                    \multicolumn{7}{l}{\hspace{1cm} \textit{marker $1$}} \\
                        \hspace{1.5cm} $\eta_{1,1}$ & 0.500 & 0.496 & -0.9 & 0.110 & 0.104 & 93.2 \\
                        \hspace{1.5cm} $\eta_{1,2}$* & 0.707 & 0.708 & 0.1 & 0.042 & 0.042 & 95.4 \\
                        \hspace{1.5cm} $\eta_{1,3}$* & 0.707 & 0.711 & 0.5 & 0.041 & 0.041 & 94.6 \\
                    \multicolumn{7}{l}{\hspace{1cm} \textit{marker $2$}} \\
                        \hspace{1.5cm} $\eta_{2,1}$ & 0.250 & 0.250 & -0.1 & 0.112 & 0.104 & 93.8 \\
                        \hspace{1.5cm} $\eta_{2,2}$* & 0.707 & 0.705 & -0.2 & 0.044 & 0.044 & 94.0 \\
                        \hspace{1.5cm} $\eta_{2,3}$* & 0.224 & 0.221 & -1.3  & 0.038 & 0.038 & 94.4 \\
                \multicolumn{7}{l}{\hspace{0.5cm} parameters for Gaussian markers} \\
                    \multicolumn{7}{l}{\hspace{1cm} \textit{marker $3$}} \\
                        \hspace{1.5cm} $\eta_{3,1}$ & 1.000 & 0.101 & 0.1 & 0.035 & 0.034 & 94.6 \\
                        \hspace{1.5cm} $\eta_{3,2}$ & 0.400 & 0.400 & -0.1 & 0.028 & 0.028 & 93.8 \\
                    \multicolumn{7}{l}{\hspace{1cm} \textit{marker $4$}} \\
                        \hspace{1.5cm} $\eta_{4,1}$ & 2.000 & 2.001 & 0.0 & 0.017 & 0.017 & 95.2 \\
                        \hspace{1.5cm} $\eta_{4,2}$ & 0.200 & 0.200 & -0.1 & 0.014 & 0.014 & 94.0 \\
                \multicolumn{7}{l}{\hspace{0.5cm} standard deviation from measurement error, $\sigma$} \\
                    \hspace{1cm} marker 1 & 1.000 & 1.006 & 0.6 & 0.084 & 0.085 & 95.2 \\
                    \hspace{1cm} marker 2 & 1.000 & 0.998 & -0.2 & 0.090 & 0.090 & 94.2 \\
                    \hspace{1cm} marker 3 & 1.000 & 1.008 & 0.8 & 0.078 & 0.077 & 93.4 \\
                    \hspace{1cm} marker 4 & 1.000 & 1.005 & 0.5 & 0.077 & 0.077 & 95.0 \\
            \hline
            \multicolumn{7}{c}{* squared root of the increment}
        \end{tabular}
    \end{center}
    \label{tab:simu}
\end{table}

%----------------------------------------------%
%--               ILLUSTRATION               --%
%----------------------------------------------%

\section{Illustration in Multi-System Atrophy}

We illustrate our methodology to the Multiple-System Atrophy (MSA), a rare neurodegenerative disease characterized by various combinations of parkinsonism, cerebellar ataxia and dysautonomic symptoms. The disease progresses very fast and is fatal with a median survival between 8 and 10 years after the first symptoms onset \cite{foubert-samier_disease_2020}. The occurrence of death suddenly interrupts the follow-up of MSA patients who are usually the most affected. This constitutes an informative dropout that needs to be accounted for in the statistical analyses to avoid biasing the model estimates. 

As many other neurodegenerative diseases, MSA clinical progression is studied almost exclusively using scales that measure the motor degradation, the dysautonomic dysfunction or the pathology-related quality-of-life. In particular, the Unified Multiple-System Atrophy Rating Scale (UMSARS) assesses disease severity in MSA patients in 4 sub-scales. The first one, reported by the patient, assesses functional impairments. 
\RVW{In this illustration, we focus essentially on one item, the 2\textsuperscript{nd} item of the first UMSARS subscale, which measures dysphagia, the difficulty to swallow food and saliva.}
Clinical assumptions are that, among the motor and functional impairments of MSA, dysphagia could represent an important aspect of the prognosis and have a substantial impact on the risk of death. This ability progressively deteriorates during the course of the disease and most severe cases require nasogastric tube or gastrostomy feeding. Dysphagia is measured as a 5-level likert scale (from 0 Normal state to 4 Severe incapacity) leading to repeated ordinal data. 

Finally, the MSA diagnosis can be made several years after the first symptoms since the first MSA symptoms may not be specific (e.g., some are similar to those a Parkinson's disease). This delayed diagnosis yields to a delayed entry in the cohort. 

By accounting for these three challenges (ordinal repeated data, informative dropout and delayed entry), our methodology is particularly useful to study MSA progression.

\subsection{Multiple-System Atrophy French cohort}

Since 2007, the university hospitals of Bordeaux and Toulouse, the two French reference centres for MSA, have constituted the MSA French cohort which includes all the patients diagnosed with a MSA, and follows them every year with a complete clinical examination including the UMSARS completion. 
For this application, we considered all the MSA cases included in the cohort who had at least one completed dysphagia item during the follow-up before the administrative censoring on December 31\textsuperscript{st} 2019. A total of 634 MSA patients were included with a total of 1819 dysphagia assessments (see description in Table \ref{tab:description}). Patients were included in the cohort 4.58 (SD=2.61) years on average after their first symptoms and were 65.1 (SD=8.2) years old in mean at entry. Two subtypes of MSA are distinguished: MSA-P for predominant parkinsonism and MSA-C for predominant cerebellar impairment. MSA-P (66.1\%) were more frequent than MSA-C in the cohort (33.9\%) as reported in Caucasian cohort \cite{wenning_natural_2013}. As for many neurodegenerative diseases, the definite diagnosis of MSA is based on post-mortem neuropathologies. In clinical practice, the MSA diagnosis is therefore based on criteria established in 2008 by Gilman et al. \cite{gilman_second_2008} and giving two degrees of diagnostic certainty: probable and possible. At first visit, 151 (23.8\%) and 483 (76.2\%) met consensus for possible MSA and probable MSA, respectively. 

Patients had in mean 2.87 (SD=2.09) clinical visits, and remained in the cohort up to 6.94 (SD=3.34) years after first symptoms onset on average. \RVW{They entered the cohort about 4.58 (SD=2.61) years after first symptoms onset.} More than half of the patients (51.89\%) died and 17.82\% were not seen in the 18 months preceding administrative censoring. As shown in Figure \ref{fig:schema_appli} (bottom right), the survival probability dropped very quickly from approximately 2 years after the first symptoms onset reaching a survival probability of 0.19 [0.16 - 0.24] after 10 years, 0.04 [0.02 - 0.06] after 15 years and 0 after 20 years. This illustrates the importance of the truncation of the follow-up by the occurrence of death in MSA. Previous analyses suggest that this truncation is related to the course of the disease \cite{foubert-samier_disease_2020}. 

Regarding dysphagia, at entry, 180 (28.39\%) patients had a normal state (no particular difficulty to swallow), 242 (38.17\%) a mild impairment (choking less than once a week), 139 (21.92\%) a moderate impairment (occasional food aspiration with choking more than once a week), 62 (9.78\%) a marked impairment (frequent food aspiration) and 6 (0.95\%) a severe state requiring assistance to be feeded.
\RVW{Note that even if the severe state is under-represented at entry, the frequency increases to $31$ during the follow-up so that it does not pose any problem of sparse category in the estimation process (in a case where a category would not be represented at all, it could be grouped with another one).} Dysphagia item degradation substantially progressed over disease time as illustrated with the individual observed trajectories (Figure \ref{fig:schema_appli} -bottom left).

\begin{table}[p]
    \caption{Description of the MSA sample at entry and over time (N=634)}
    \begin{center}
        \begin{tabular}{lcc} 
            \hline
            Characteristic & N (\%) & mean $\pm$ sd \\
            \hline
            At entry \\
                \hspace{0.5cm} Sex \\
                    \hspace{1cm} Male & 311 (49.05\%) & \\
                    \hspace{1cm} Female & 323 (50.95\%) & \\
                \hspace{0.5cm} Center \\
                    \hspace{1cm} Bordeaux & 320 (50.47\%) & \\
                    \hspace{1cm} Toulouse & 314 (49.53\%) & \\
                \hspace{0.5cm} Diagnosis \\
                    \hspace{1cm} MSA-C, with predominant cerebellar impairment & 215 (33.91\%) & \\
                    \hspace{1cm} MSA-P, with predominant parkinsonism & 419 (66.09\%) & \\
                \hspace{0.5cm} Diagnosis certainty \\
                    \hspace{1cm} Possible & 151 (23.82\%) & \\
                    \hspace{1cm} Probable & 483 (76.18\%) & \\
                \hspace{0.5cm} Age at first symptoms onset & & 60.48 $\pm$ 8.32  \\
                \hspace{0.5cm} Age at entry & & 65.06 $\pm$ 8.19 \\
                \hspace{0.5cm} Years since first symptoms onset & & 4.58 $\pm$ 2.61 \\
                \hspace{0.5cm} Dysphagia item & & 1.16 $\pm$ 0.98 \\
                    \hspace{1cm} 0. Normal & 180 (28.39\%) & \\
                    \hspace{1cm} 1. Mild impairment & 242 (38.17\%) & \\
                    \hspace{1cm} 2. Moderate impairment & 139 (21.92\%) & \\
                    \hspace{1cm} 3. Marked impairment & 62 (9.78\%) & \\
                    \hspace{1cm} 4. Nasogastric tube or gastrostomy feeding & 6 (0.95\%) & \\
            \hline
            During follow-up \\
                \hspace{0.5cm} Visits with dysphagia item completed & 1819 (98\%) & \\
                \hspace{0.5cm} Dysphagia item observations per patient & & 2.87 $\pm$ 2.09 \\
                \hspace{0.5cm} Years of follow-up & & 6.94 $\pm$ 3.34 \\
                \hspace{0.5cm} Patients with nasogastric tube or gastrostomy feeding & 31 (4.89\%) & \\
                \hspace{0.5cm} Drop-out & 113 (17.82\%) & \\
                \hspace{0.5cm} Death & 329 (51.89\%) & \\
            \hline
        \end{tabular}
    \end{center}
    \label{tab:description}
\end{table}

\subsection{Specification of the model} \label{sec:MSAmodelspec}

\begin{figure}[p]
\caption{Schematic diagram of the shared random effect joint model applied to MSA dysphagia study, along with observed individual observed trajectory of dysphagia, and Kaplan-Meier estimate of survival probability (N=634)}
\centering
\includegraphics[width = 14cm, trim= 0cm 0cm 8.5cm 0cm, clip, page = 2]{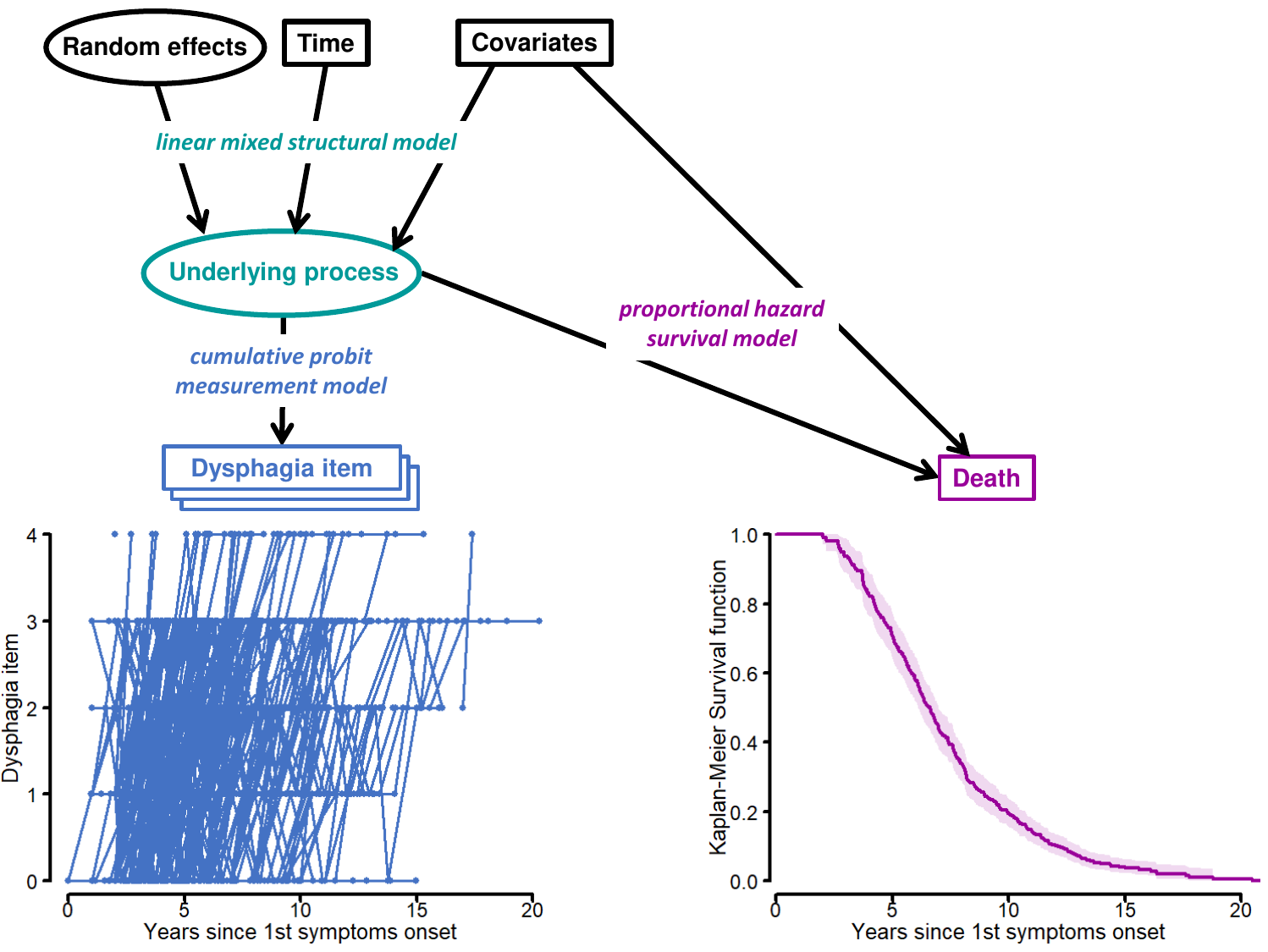}
\label{fig:schema_appli}
\end{figure}

This study aimed at better understanding progression of swallowing difficulties over disease time and at quantifying its association with death. To do so, we built a shared random effect joint model as illustrated in Figure \ref{fig:schema_appli}, and considered time since first symptoms as the study timescale to describe the natural history of the item progression. \RVW{The model handles delayed entry due to the fact that patients may enter the study years after the first symptoms occurred}.

The observed 5-level item probability was linked to the underlying continuous dysphagia process at the exact same time by a cumulative probit model (cf. Equation \eqref{eq:mks-D}). The trajectory of the underlying dysphagia process was simultaneously described over time by a latent linear mixed model (cf. Equation \eqref{eq:Delta}) \RVW{with individual random effects on each function of time considered to take into account the within patient correlation. After preliminary analyses comparing a linear trajectory, a quadratic trajectory and a trajectory approximated by natural cubic splines with 1 internal knot, we selected the quadratic trajectory as the one providing the best fit to the data according to the Akaike Information Criterion (AIC) (AIC=5944.53, AIC=5902.23, AIC=6021.21 for the models with linear, quadratic and splines trajectories, respectively).} The association between dysphagia and death was simultaneously modelled using a survival model where the instantaneous risk of death was a function of the current underlying level of dysphagia (cf. Equation \eqref{eq:surv} with option \eqref{eq:surv_CL}). The baseline hazard function was defined by cubic M-splines with \RVW{3 internal knots placed at the quantiles of the observed times of event, that is $4.65$, $6.37$, $8.71$ years. The boundary knots were placed at $0$ and $24$ years.} The structural longitudinal model and the survival model were adjusted for sex (male or female), diagnosis (MSA-C or MSA-P), certainty of the diagnosis (possible or probable) and age at the first symptoms onset. 

\subsection{Results}

\begin{figure}[p]
\caption{Mean trajectories of dysphagia item predicted by the shared random effect joint model according to the four main covariates (sex, age at first symptoms onset, MSA subtype and certainty degree of diagnosis). The reference profile is a male patient, 60 years old at first symptoms onset, diagnosed MSA-C with possible certainty. Shades represent the 95\% confidence intervals obtained by Monte Carlo approximation with 1000 draws. The reported p-values are those of Wald tests for the association at inclusion and the association with the functions of time.}
    \centering
    \includegraphics[width=15cm]{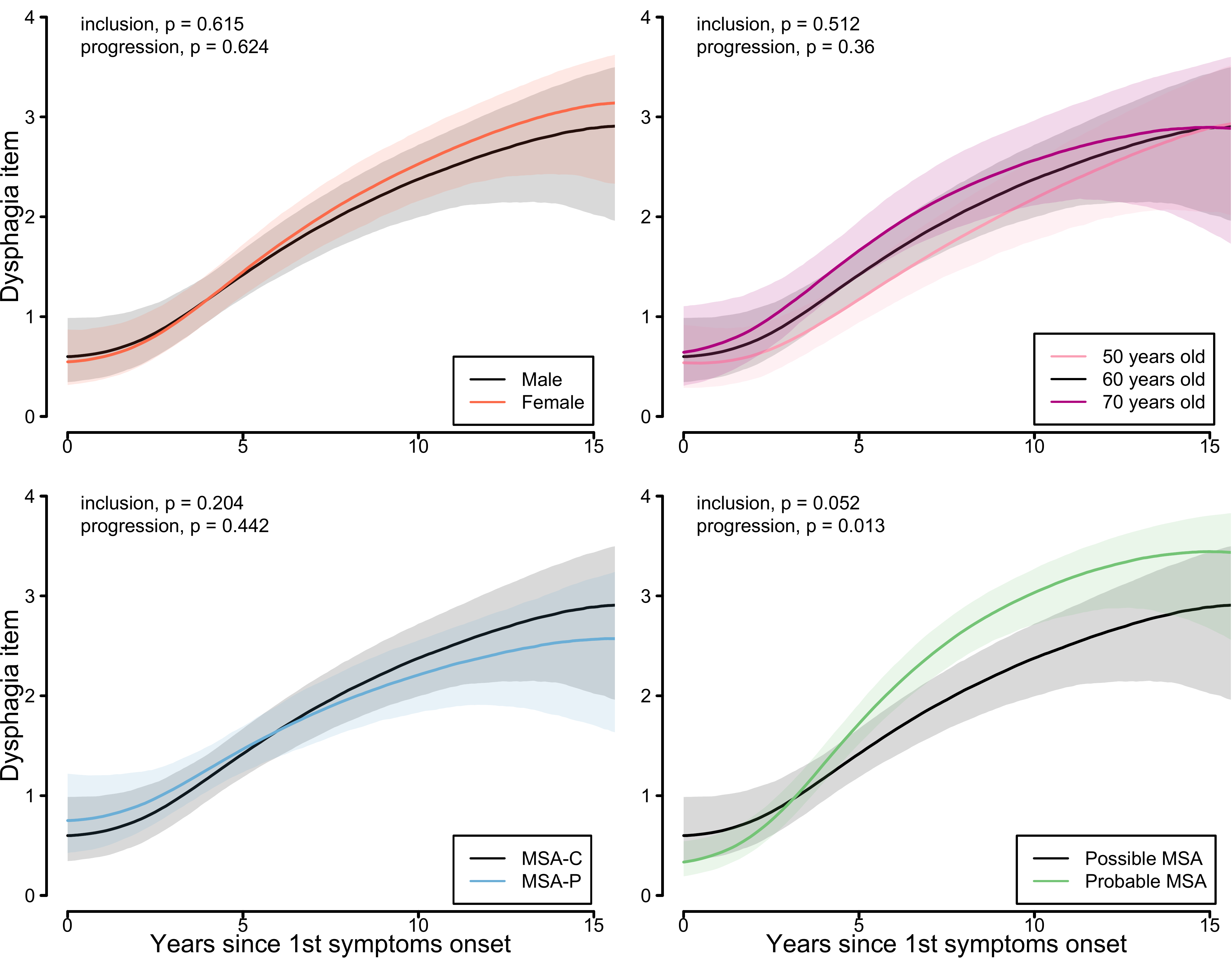}
    \label{fig:traj}
\end{figure}

The estimates of the joint model are provided in supplementary table. The predicted trajectories of dysphagia item are displayed in Figure \ref{fig:traj} according to profiles of patient differing by the four main covariates of interest (sex, age at first symptoms onset, MSA subtype and certainty degree of diagnosis). The reference profile corresponded to a male patient, 60 years old at first symptoms onset, diagnosed MSA-C with possible certainty (black line). 

Overall, the trajectories of dysphagia did not significantly differ according to sex, age and MSA subtype. Male and female patients seemed to endure the same dysphagia progression over at least the first 5 years of the disease. Female patients seemed to face a more important degradation afterwards although the dysphagia progression was not statistically different by sex (interaction on time and quadratic time, p=0.624). Dysphagia evolved slighly faster for older patients with substantial differences during the follow-up. At 5 years of disease, the dysphagia level was 1.662 (95\%CI=1.400,1.956), 1.419 (95\%CI=1.184,1.671) and 1.173 (95\%CI=0.943,1.445) for patients aged 70, 60 and 50 at the beginning of the disease, respectively. This highlights an approximate gap of 0.5 points between patients with 20 years difference, stable between 5 and 10 years of disease.

No difference in dysphagia progression was observed between MSA-C and MSA-P patients. At the beginning of the disease, MSA-P patients seemed to have a little higher level of dysphagia than MSA-C patients, but after 6 years of disease, this trend reversed and MSA-C patients surpassed MSA-P patients in terms of degradation degree. This could be explained by the fact that MSA-P patients are initially more affected on the motor level as opposed to MSA-C patients who suffer from non-motor and dysautonomia symptoms \cite{foubert-samier_disease_2020}.

Trajectories of dysphagia substantially differed according to the diagnosis certainty. Patients diagnosed with possible MSA had a significantly higher level of dysphagia at the beginning of the disease compared to patients diagnosed with more confidence (probable) with a mean level of 0.600 (95\%CI=0.344,0.986) and 0.334 (95\%CI=0.192,0.542) at onset for possible and probable, respectively. However, patients with probable MSA progressed much more rapidly reaching a mean level of 3.034 (95\%CI= 2.711,3.272) after only 10 years while possible MSA patients still had in mean 2.377 (95\%CI=1.995,2.722). 

The trajectory of dysphagia was also substantially associated with the risk of death through its current underlying level (HR = 3.773, 95\%CI =2.588,5.501) after adjustment on sex, age at first symptoms onset, MSA subtype and certainty degree of diagnosis (Table \ref{tab:surv}\RVW{, right column}). This means that the level of dysphagia at a certain time $t$ is strongly associated with the instantaneous risk of death at the exact same time: the higher the level of dysphagia, the higher the risk of death. \RVW{A second model was estimated without accounting for dysphagia progression (i.e., fixing its association parameter to $\alpha=0$) to assess whether some covariate association with death were changed when considering the association between dysphagia and death. Results are reported in Table \ref{tab:surv} (left-column)}. 
\RVW{The association with age at first symptoms was not impacted by the adjustment on dysphagia (HR = 1.215, 95\%CI = 1.063,1.388 for a 10-year difference and HR = 1.160, 95\%CI = 1.007,1.337 for the model without and with dysphagia, respectively). This suggests} that the effect of age on the risk of death in this pathology may be independent of the level of clinical progression. Older patients at first symptoms have a higher risk of death.
\RVW{As well,} with or without adjustment for dysphagia, there was no evidence of difference of death risk according to MSA \RVW{diagnosis} subtype \RVW{(HR = 1.081,  95\%CI = 0.857,1.363 without dysphagia, HR = 1.132,  95\%CI = 0.889,1.441 with dysphagia)}.
Furthermore, the higher risk of death of probable cases was slightly attenuated when accounting for the level of dysphagia impairment \RVW{(from HR = 1.864,  95\%CI = 1.398,2.485 without dysphagia to HR = 1.376,  95\%CI = 1.026,1.844) with adjustment for}. This was expected as we saw that probable patients had a significantly higher level of dysphagia than possible patients (Figure \ref{fig:traj}).
\RVW{Finally, the association between sex and risk of death was not substantially changed when adjusting for the underlying level of dysphagia (HR = 0.887,  95\%CI = 0.713,1.103 without adjustment to HR = 0.827,  95\%CI = 0.660,1.036 with adjustment for dysphagia).}

\begin{table}[p]
    \caption{Hazard ratios for the risk of death in proportional hazard models when accounting or not for the progression of dysphagia over time (N=634)}
    \begin{center}
        \begin{tabular}{lcc}
            \hline
             & \multicolumn{2}{c}{Hazard ratios [95\%CI]} \\
            \cline{2-3}
            Covariate, modality (reference modality) & Without dysphagia & With dysphagia \\
            \hline
            Sex, &  &      \\
            \hspace{0.3cm} Women (ref Men) & 0.887 [ 0.713 ; 1.103 ] & 0.827 [ 0.660 ; 1.036 ]     \\
            Age at first symptoms onset, &  &      \\
            \hspace{0.3cm} 10 years gap (ref 60 years old) & \textbf{1.215 [ 1.063 ; 1.388 ]} & \textbf{1.160 [ 1.007 ; 1.337 ]}     \\
            Diagnosis, &  &       \\
            \hspace{0.3cm} MSA-P (ref MSA-C) & 1.081 [ 0.857 ; 1.363 ] & 1.132 [ 0.889 ; 1.441 ]      \\
            Diagnosis certainty, &  &        \\
            \hspace{0.3cm} Probable MSA (ref Possible MSA) & \textbf{1.864 [ 1.398 ; 2.485 ]} & \textbf{1.376 [ 1.026 ; 1.844 ]}       \\
            Current level of dysphagia, &  &       \\
            \hspace{0.3cm} unit = 1 SD at first symptoms & -  & \textbf{3.773 [ 2.588 ; 5.501 ]}      \\
            \hline
        \end{tabular}
    \end{center}
    \label{tab:surv}
\end{table}

\subsubsection{\RVW{Goodness-of-fit assessment}}

\RVW{The goodness-of-fit of the final model was investigated using two graphical tools:}

\begin{itemize}
\item \RVW{for the survival submodel, we generated 500 samples from the fitted joint model and compared the distribution of the
 corresponding predicted survival curves with the one observed in the original sample. As shown in Figure S1 in supplementary materials, the survival curves predicted by the model were very close to the observed one, showing how the joint model adequately fitted the time-to-event data. }
 
\item \RVW{for the longitudinal submodel, we computed the predicted item values derived from the estimated joint model, and compared the mean trajectory of these predictions with the mean trajectory of the observations by splitting time into  1-year intervals. The predicted item values were obtained by (i) predicting the patient-specific vector of random effect as the mode of the posterior distribution of the random effects given the patient observations, and (ii) computing the expected item level according to the predicted patient-specific random-effect. As shown in Figure S2 in supplementary materials, the predicted item mean trajectory was close to the observations, illustrating the adequacy of the estimated model to the longitudinal data. }
\end{itemize}

%-------------------------------------------%
%--               DISCUSSION              --%
%-------------------------------------------%

\section{Concluding remarks}

The joint modelling of longitudinal data and time-to-event data has become a standard methodology to address the problem of informative dropout and more generally investigate the association between longitudinal markers and clinical events \cite{rizopoulos_introduction_2014}. In this work, we have shown how this methodology could be extended to longitudinal data of different natures by separating the longitudinal model for the quantity of interest defined as a latent process from the measurement model that links the quantity of interest to its repeated observations. This model thus naturally handles univariate or multivariate markers of the same underlying quantity but also continuous (Gaussian or not) or discrete markers. Specific examples include the case of a unique continuous Gaussian marker which corresponds to the classical joint model framework \cite{rizopoulos_joint_2012} or the case of various ordinal items measuring the same underlying construct. In the latter case, the longitudinal submodel is a dynamic Item Response Theory (IRT) submodel (see \cite{proust-lima_continuous-time_2021,barbieri_methodes_2016} in this special issue). \RVW{Although not of importance in the application to MSA, the methodology also handles competing risks of events, a situation which is often encountered in cohorts when different types of clinical progression and/or dropout may occur (e.g., types of recurrence in cancer research, multiple causes of death).}

All joint models assume a given structure of correlation between the longitudinal data and the time-to-event data. In the literature, two main structure types can be found, \RVW{the \it{shared random-effect models}} which consider that the random effects from the longitudinal model capture all the correlation between the longitudinal process and the time-to-event process, and the \RVW{\it{joint latent class models}} which assume that a latent group structure captures all the correlation between the two processes \cite{proust-lima_joint_2014}. In this work, we \RVW{chose} to focus on the extension of the shared random effect joint model \RVW{which enables an explicit quantification of the association with hazard ratios, as illustrated in MSA, and which is more often used in practice, and has been usually favored for treating informative dropout \cite{thomadakis_longitudinal_2019,dantan_pattern_2008}. We refer to Proust-Lima et al. \cite{proust-lima_joint_2014} and R package lcmm \cite{proust-lima_estimation_2017} for further details on joint latent class models incorporating multivariate outcomes measuring the same underlying process.} 

Within the framework of shared random effect joint models, we considered a dependence structure on the current level of the latent process or on the individual random-effects of the structural longitudinal model. However, other dependence structures could also be worth exploring such as the current slope of the latent process, an accumulation of the latent process over time \cite{andrinopoulou_bayesian_2016} or nonlinear functions of them \cite{sene_shared_2013}. The choice of the dependence structure depends on the application framework and is part of the model specification and building. \RVW{The R-package JLPM is still under development to provide more options for the dependency structure between the two processes.}

\RVW{For simplicity, we described the methodology of the latent process model under the assumption of measurement invariance. A lack of measurement invariance can occur when markers present a different functioning according to a specific covariate (called Differential Item Functioning in IRT literature) or according to time (called response shift in IRT literature). Such lack of invariance can be explored with our methodology and is implemented in the software solution. We refer to another paper in this special issue for further details \cite{proust-lima_modeling_2022}.} 

To conclude, measurement scales and questionnaires become more and more central in health studies to measure the physical, mental or psychological state of patients. With this joint model implemented in the R-package JLPM, we provide a new and relevant solution to study their trajectories over time and their association with events of interest. 

\vspace{1cm}
\noindent \textbf{Funding} This work was funded by the French National Research Agency
455 (Project DyMES - ANR-18-C36-0004-01).

%~~~~~~~~~~~~~~~~~~~~~~~~%
%~~     REFERENCES     ~~%
%~~~~~~~~~~~~~~~~~~~~~~~~%

%\newpage

\bibliographystyle{elsarticle-num}
\bibliography{main_arXiv}

%~~~~~~~~~~~~~~~~~~~~~~~~~~~~~~~~~~%
%~~    SUPPLEMENTARY MATERIAL    ~~%
%~~~~~~~~~~~~~~~~~~~~~~~~~~~~~~~~~~%

\includepdf[pages={1-}]{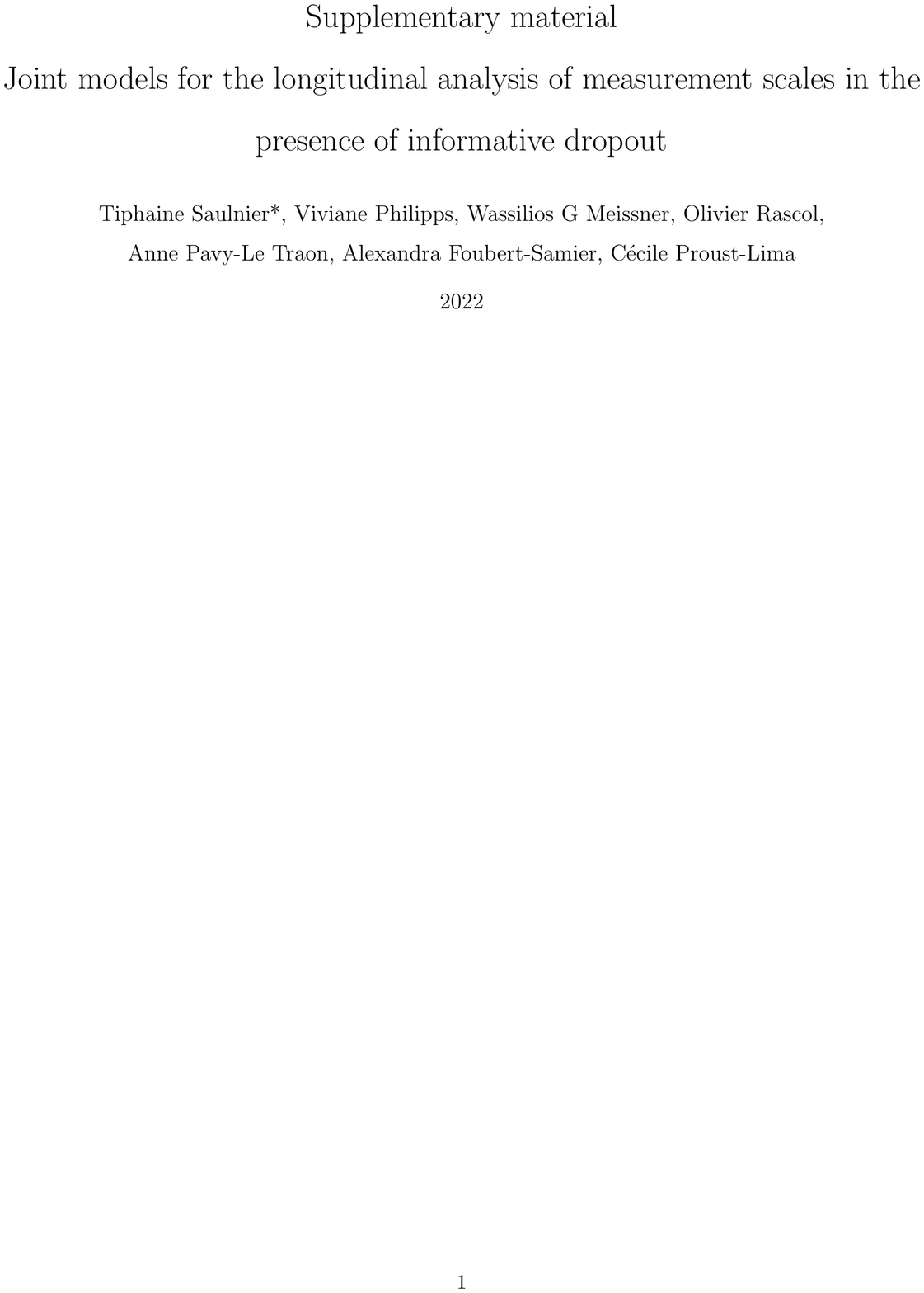}

\end{document}